\documentclass[12pt]{article}                                                   
\usepackage{amsmath}                                                            
\usepackage{txfonts}                                                            
\usepackage{amssymb}                                                            
\usepackage{amsfonts}                                                           
\usepackage{graphicx}                                                           
\usepackage{latexsym}                                                           
\usepackage[dvips]{color}                                                       
\usepackage{epsfig}                                                             
\usepackage{wasysym}                                                            
\def\3{\ss}

\def\kreis{\raise0.85pt\hbox{$\scriptstyle\bigcirc$}}
\def\vollk{\lower0.85pt\hbox{\Large $\bullet$}}

\begin{document}

\title{
\vspace{-2.5cm}
\flushleft{\normalsize DESY 10-014} \\
\vspace{-0.35cm}
{\normalsize Edinburgh 2010/03} \\
\vspace{-0.35cm}
{\normalsize Liverpool LTH 861} \\
\vspace{-0.35cm}
{\normalsize February 2010} \\
\vspace{0.5cm}
\centering{\Large \bf Pion in a Box}}
%
%
\author{W. Bietenholz$^a$, M. G\"ockeler$^b$, R. Horsley$^c$,
  Y. Nakamura$^b$, D. Pleiter$^d$,\\ 
  P.E.L. Rakow$^e$, G. Schierholz$^{f,b}$ and J.M. Zanotti$^c$\\[1.00em]
        $^a$ Insituto de Ciencias Nucleares, Universidad Nacional
        Aut\'{o}noma de M\'{e}xico\\ 
        A.P. 70-543, C.P. 04510 Distrito Federal, Mexico\\[0.25em]
        $^b$ Institut f\"ur Theoretische Physik, Universit\"at Regensburg,\\
        93040 Regensburg, Germany\\[0.25em]
        $^c$ School of Physics and Astronomy, University of Edinburgh,\\
        Edinburgh EH9 3JZ, UK\\[0.25em]
        $^d$ NIC/DESY, 15738 Zeuthen, Germany\\[0.25em]
        $^e$ Theoretical Physics Division, Department of Mathematical
        Sciences,\\ University of Liverpool, Liverpool L69 3BX, UK\\[0.25em]
        $^f$ Deutsches Elektronen-Synchrotron DESY,\\
        22603 Hamburg, Germany\\[1.0em]
        -- QCDSF Collaboration --}

\date{ }

\maketitle

\begin{abstract}
\noindent
The residual mass of the pion in a finite spatial box at vanishing
quark masses, the mass gap, is computed with two flavors of
dynamical clover fermions. The result is compared with predictions of chiral
perturbation theory in the $\delta$ regime.  
\end{abstract}

Understanding finite size effects has always been an important issue in
lattice studies of QCD. For an accurate determination of hadron
observables it is of foremost importance to account for finite volume
corrections. In addition, finite size effects can provide valuable
information on the low-energy physics of the system.
As lattice calculations are reaching the physical pion mass,
the impact of finite volume will become pronounced. The main effect is
caused by modified pion dynamics arising from the boundary conditions being
imposed on the system.

In a finite spatial box chiral symmetry does not break down
spontaneously. This results in a mass gap, even if the quark masses
vanish. The physics behind that is described by a
simple quantum mechanical rotator~\cite{Leutwyler:1987ak}, whose 
energy levels can be computed from an expansion of the chiral
effective theory in the $\delta$ 
regime $m_\pi L \ll 1, T \gg L$, where $L$ ($T$) is the spatial
(temporal) extent of the box. We
are now in the position to probe the pion mass near the chiral
limit and account for this effect. What makes this problem attractive,
beyond the computational 
task of solving QCD in a finite volume, is that it is 
a universal feature of quantum mechanics, which is only constrained by
the symmetry of the system. For a previous attempt of extracting the
mass gap see~\cite{Hasenfratz:2006xi}.

In a series of papers Leutwyler~\cite{Leutwyler:1987ak}, Hasenfratz and
Niedermayer~\cite{H&N} and Hasenfratz~\cite{Hasenfratz:2009mp} have
computed the energy levels for two flavors of dynamical
quarks to leading (L), next-to-leading (NL) and
next-to-next-to-leading (NNL) order. The mass gap, {\it i.e.} the
residual mass of the pion in the chiral limit, up to NNL order turns out to be
\begin{equation}
m_\pi^{\rm res} = \frac{3}{2 F_\pi^2 L^3 (1+\Delta)}
\label{rotor}
\end{equation}
with
\begin{equation}
\begin{split}
\Delta &= \frac{2}{F_\pi^2 L^2}\,0.2257849591 \\
       &+ \frac{1}{F_\pi^4 L^4}\,
\left[0.088431628 - \frac{0.8375369106}{3\pi^2}\Bigg(\frac{1}{4}
  \ln{(\Lambda_1 L)^2} +  \ln{(\Lambda_2 L)^2}\Bigg)\right]\, ,
\end{split}
\end{equation}
where $F_\pi$ is the pion decay constant, and $\Lambda_i$
are the intrinsic scale parameters of the low-energy
constants~\cite{Colangelo:2001df}, 
\begin{equation}
\bar{l}_i = \ln\left(\frac{\Lambda_i}{m_\pi^{\rm phys}}\right)^2\, .
\end{equation}

Our simulations are performed with the Wilson gauge action and two
flavors of mass-degenerate $O(a)$ improved Wilson fermions on a range
of lattice volumes with periodic (antiperiodic) boundary conditions in
the spatial (temporal) direction. Two values of the gauge coupling,
$\beta=5.29$ and 
$5.40$, have been selected here. The hopping parameters and the
lattice volumes of the individual runs are listed in
Table~\ref{par}. It is convenient to express the scale in terms of the
force parameter $r_0$. In the chiral limit we find $r_0/a=6.201(25)$
and $6.946(44)$, respectively. We set the scale by the nucleon mass, which
gives $r_0=0.467 \, \mbox{fm}$~\cite{Gockeler:2009pe}. This
translates into lattice spacings $a=0.075$ and $0.067\, \mbox{fm}$. 

\begin{table}[t]
\begin{center}
\begin{tabular}{|c|c|c|}\hline
$\beta$ & $\kappa_{\rm sea}$ & $V\, [a^4]$ \\
\hline
 5.29 & 0.13400 & $16^3\times 32$ \\ \hline
 5.29 & 0.13500 & $16^3\times 32$ \\ \hline
 5.29 & 0.13550 & $16^3\times 32$ \\ 
 5.29 & 0.13550 & $24^3\times 48$ \\ \hline 
 5.29 & 0.13590 & $16^3\times 32$ \\ 
 5.29 & 0.13590 & $24^3\times 48$ \\ \hline
 5.29 & 0.13620 & $24^3\times 48$ \\ \hline
 5.29 & 0.13632 & $24^3\times 48$ \\
 5.29 & 0.13632 & $32^3\times 64$ \\
 5.29 & 0.13632 & $40^3\times 64$ \\ \hline
 5.29 & 0.13640 & $40^3\times 64$ \\
\hline
 5.40 & 0.13500 & $24^3\times 48$ \\ \hline
 5.40 & 0.13560 & $24^3\times 48$ \\ \hline
 5.40 & 0.13610 & $24^3\times 48$ \\ \hline
 5.40 & 0.13625 & $24^3\times 48$ \\ \hline
 5.40 & 0.13640 & $24^3\times 48$ \\
 5.40 & 0.13640 & $32^3\times 64$ \\ \hline
 5.40 & 0.13660 & $32^3\times 64$ \\ \hline
\end{tabular}
\end{center}
\caption{Parameters of the lattice data sets used in this analysis.}  
\label{par}
\end{table}

The quark masses are computed from the ratio (see {\it e.g.}~\cite{Gockeler:2006jt}, and references therein)
\begin{equation}
am_q = \frac{\langle a\partial \mathcal{A}_4(t) P(0)\rangle}{\langle
  P(t) P(0)\rangle} 
\end{equation} 
at $t \gg a$, where $\mathcal{A}_\mu$ is the improved axial vector
current~\cite{Della Morte:2005se},  
\begin{equation}
\mathcal{A}_\mu = A_\mu + c_A\, a\partial_\mu P \, ,
\end{equation}
and $A_\mu$ and $P$ are the standard axial vector current and 
pseudoscalar density, respectively. The quark masses show practically no
finite size effects beyond the statistical errors. In Table~\ref{tab}
we give the results from the largest lattices. Unlike the quark
masses, the pion masses show significant finite size effects. The
results obtained on the various volumes are shown in Table~\ref{tab} as
well. The pion masses from the lattices in Table~\ref{par} are printed
in roman. They reach as low as $170\, \mbox{MeV}$. 

\begin{table}[t]
\begin{center}
\begin{tabular}{|c|c|l|l|l|l|l|}\hline
$\beta$ & $\kappa_{\rm sea}$ & \multicolumn{4}{|c|}{$am_\pi$} &
  \multicolumn{1}{|c|}{$am_q$} \\ 
 &  & \multicolumn{1}{|c|}{$L/a=16$} & \multicolumn{1}{|c|}{$L/a=24$}
  & \multicolumn{1}{|c|}{$L/a=32$} & \multicolumn{1}{|c|}{$L/a=40$} &
  \\ 
\hline
 5.29 & 0.13400 & 0.5767(11) & & & & 0.08781(18) \\
 5.29 & 0.13500 & 0.4206(9)  &  {\it 0.4195(9)} & & & 0.05113(10) \\
 5.29 & 0.13550 & 0.3325(13) &  0.3270(6) & {\it 0.3268(6)} & {\it
   0.3268(6)} & 0.03293(8) \\
 5.29 & 0.13590 & 0.2518(15) &  0.2395(5) & {\it 0.2388(5)} & {\it
   0.2388(5)} & 0.01873(4) \\ 
 5.29 & 0.13620 & & 0.1552(6) & {\it 0.1534(6)} & {\it 0.1532(6)} &
 0.00792(4) \\ 
 5.29 & 0.13632 & & 0.1106(12)& 0.1075(9) & 0.1034(8) & 0.00369(3) \\
 5.29 & 0.13640 & & & & 0.066(1) & 0.00137(2)
 \\
\hline
 5.40 & 0.13500 & & 0.4030(4)  & {\it 0.4030(4)} & & 0.05641(5)  \\
 5.40 & 0.13560 & & 0.3123(7)  & {\it 0.3121(7)} & & 0.03612(5)  \\
 5.40 & 0.13610 & & 0.2208(7)  & {\it 0.2200(7)} & & 0.01930(4)  \\
 5.40 & 0.13625 & & 0.1902(6)  & {\it 0.1889(6)} & & 0.01431(3)  \\
 5.40 & 0.13640 & & 0.1538(10) & 0.1504(4) & & 0.00932(3)  \\
 5.40 & 0.13660 & & {\it 0.0947(11)} & 0.0867(11) & & 0.00274(4) \\\hline
\end{tabular}
\end{center}
\caption{The pion and bare quark
  masses on the various lattices. The roman numbers are from our
  simulations on the quoted volumes, while the italic numbers have
  been obtained by extrapolation as described in the text.}  
\label{tab}
\end{table}

Some of the pion masses at the larger quark masses are only known on
$16^3$ and $24^3$ lattices. Furthermore, we miss one pion mass at the
smallest quark 
mass on the $24^3$ lattice at $\beta=5.40$. In these cases the pion
masses have been extrapolated to the 
`missing' volumes by means of the $O(p^4)$ finite size shift
formula~\cite{Colangelo:2005gd} 
\begin{equation} 
\frac{m_\pi(L)-m_\pi}{m_\pi} = -\sum_{\vec{n} \neq 0} 
\frac{x}{2 \lambda} \left[I_{m_{PS}}^{(2)}(\lambda) + x
  I_{m_{PS}}^{(4)}(\lambda)\right] \, ,
\label{fs}
\end{equation}
valid in the $p$ regime $m_\pi L \gg 1$, with
\begin{equation}
x=\frac{m_\pi^2}{16\pi^2 F_\pi^2} \, , \quad \lambda = m_\pi |\vec{n}|\,
L \, ,
\end{equation}
where
\begin{equation}
\begin{split}
I_{m_{PS}}^{(2)}(\lambda) &= -B^0(\lambda)\, , \\
I_{m_{PS}}^{(4)}(\lambda) &= \left(-\frac{55}{18} + 4\bar{l}_1 +
  \frac{8}{3}\bar{l}_2 - \frac{5}{2}\bar{l}_3-2\bar{l}_4\right)B^0(\lambda)\\[0.5em]
 &+ \,\left(\frac{112}{9} - \frac{8}{3}\bar{l}_1 -
   \frac{32}{3}\bar{l}_2\right)B^2(\lambda)
  + S_{m_{PS}}^{(4)}(\lambda) \, ,\\
 S_{m_{PS}}^{(4)}(\lambda) &= \frac{13}{3} g_0 B^0(\lambda)-\frac{1}{3}\left(40 g_0 + 32
   g_1 + 26 g_2\right)B^2(\lambda) 
\end{split}
\end{equation}
with
\begin{equation}
B^0(\lambda)= 2 K_1(\lambda) \,, \quad B^2(\lambda) = 2 K_2(\lambda)/\lambda \, .
\end{equation}
We take $F_\pi=92.4\,\mbox{MeV}$ throughout this paper. The Taylor
coefficients $g_i$ are 
\begin{equation}
g_0=2-\frac{\pi}{2} \, , \quad g_1=\frac{\pi}{2}-\frac{1}{2} \, ,
\quad g_2=\frac{1}{2}-\frac{\pi}{8} \, .
\end{equation}
The first (second) expression in the square brackets of eq.~(\ref{fs})
corresponds to $O(p^2)$\, $(O(p^4))$ corrections in the Lagrangian. 
The low-energy parameters $\bar{l}_1$, $\bar{l}_2$ and $\bar{l}_4$ are
taken from~\cite{Colangelo:2001df}: 
\begin{equation}
\bar{l}_1\simeq -0.4 \, , \quad \bar{l}_2\simeq 4.3 \, , \quad
\bar{l}_4\simeq 4.4 \, ,
\label{lec}
\end{equation}

\clearpage
\begin{figure}
\vspace*{-2.75cm}
\begin{center}
\epsfig{file=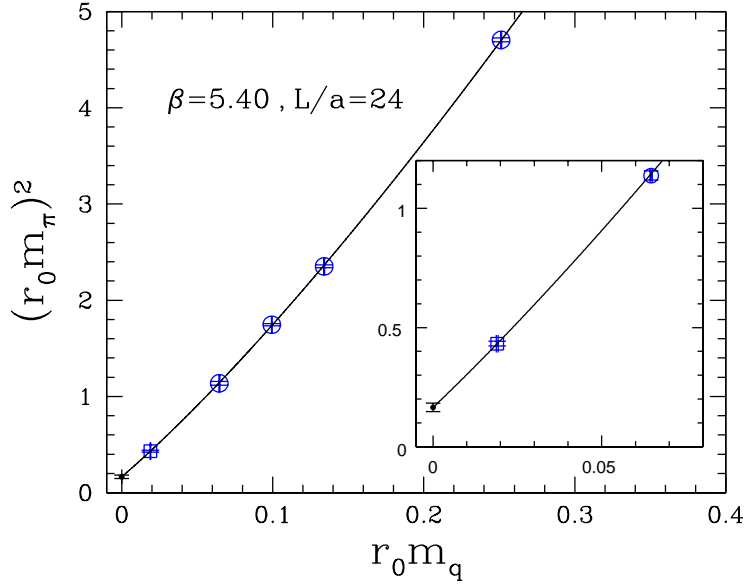,width=10cm,clip=}
\end{center}
\vspace*{-0.4cm}
\caption{The pion mass squared against the quark mass on the $24^3$
  lattice at $\beta=5.40$, together with the chiral fit
  (\ref{ch}). The circles refer to the roman numbers in
  Table~\ref{tab}, obtained directly on the quoted lattice, while the
  square is the extrapolated value and 
  refers to the italic number in Table~\ref{tab}.}
\label{fig24a}
\end{figure}
%
\begin{figure}
\vspace*{-2.75cm}
\begin{center}
\epsfig{file=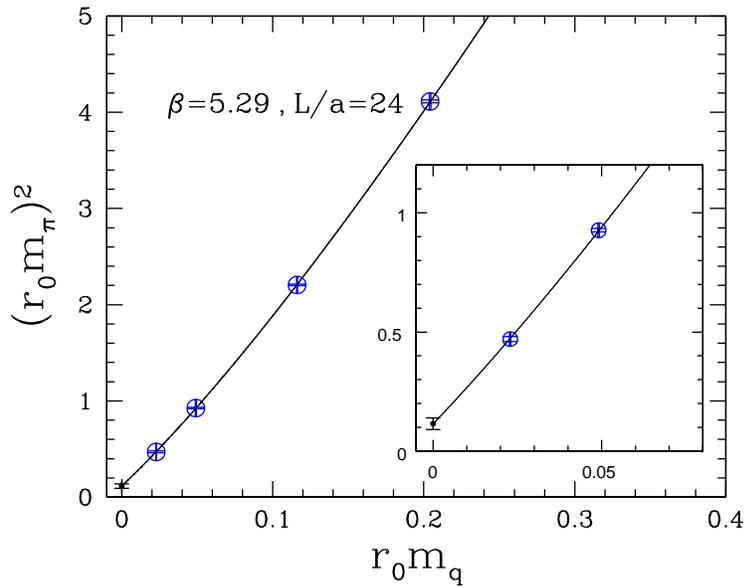,width=10cm,clip=}
\end{center}
\vspace*{-0.4cm}
\caption{The pion mass squared against the quark mass on the $24^3$
  lattice at $\beta=5.29$, together with the chiral fit
  (\ref{ch}). The circles refer to the roman numbers in 
  Table~\ref{tab}, obtained directly on the quoted lattice. The
  extrapolated value referring to the italic number in Table~\ref{tab}
  is not shown here.}
\label{fig24b}
\end{figure}

\clearpage
\begin{figure}
\vspace*{-2.75cm}
\begin{center}
\epsfig{file=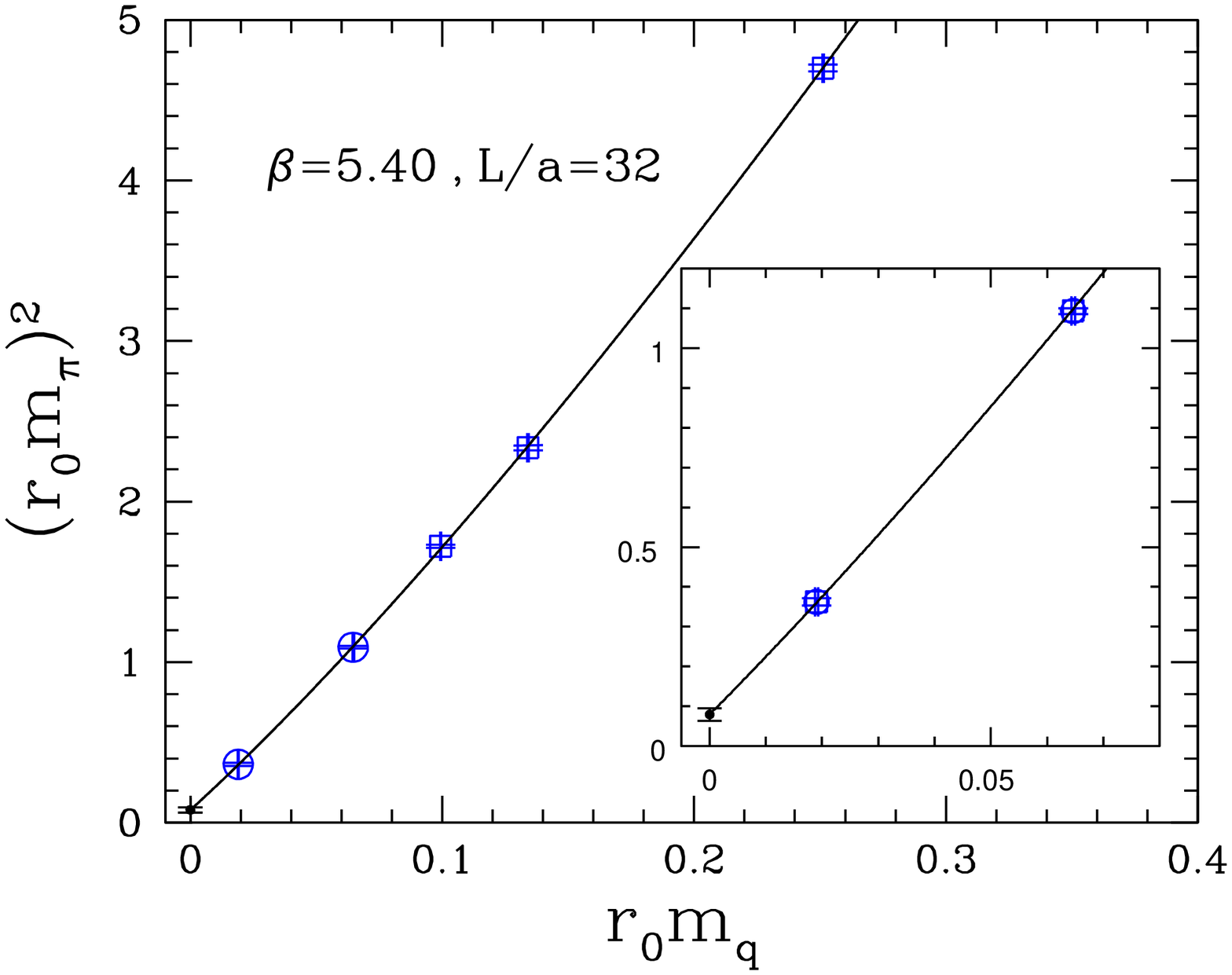,width=10cm,clip=}
\end{center}
\vspace*{-0.4cm}
\caption{The pion mass squared against the quark mass on the $32^3$
  lattice at $\beta=5.40$, together with the chiral fit
  (\ref{ch}). The circles refer to the roman numbers in 
  Table~\ref{tab}, obtained directly on the quoted lattice, while the
  squares are the extrapolated values and 
  refer to the italic numbers in Table~\ref{tab}.}
\label{fig32}
\end{figure}
%
\begin{figure}
\vspace*{-2.75cm}
\begin{center}
\epsfig{file=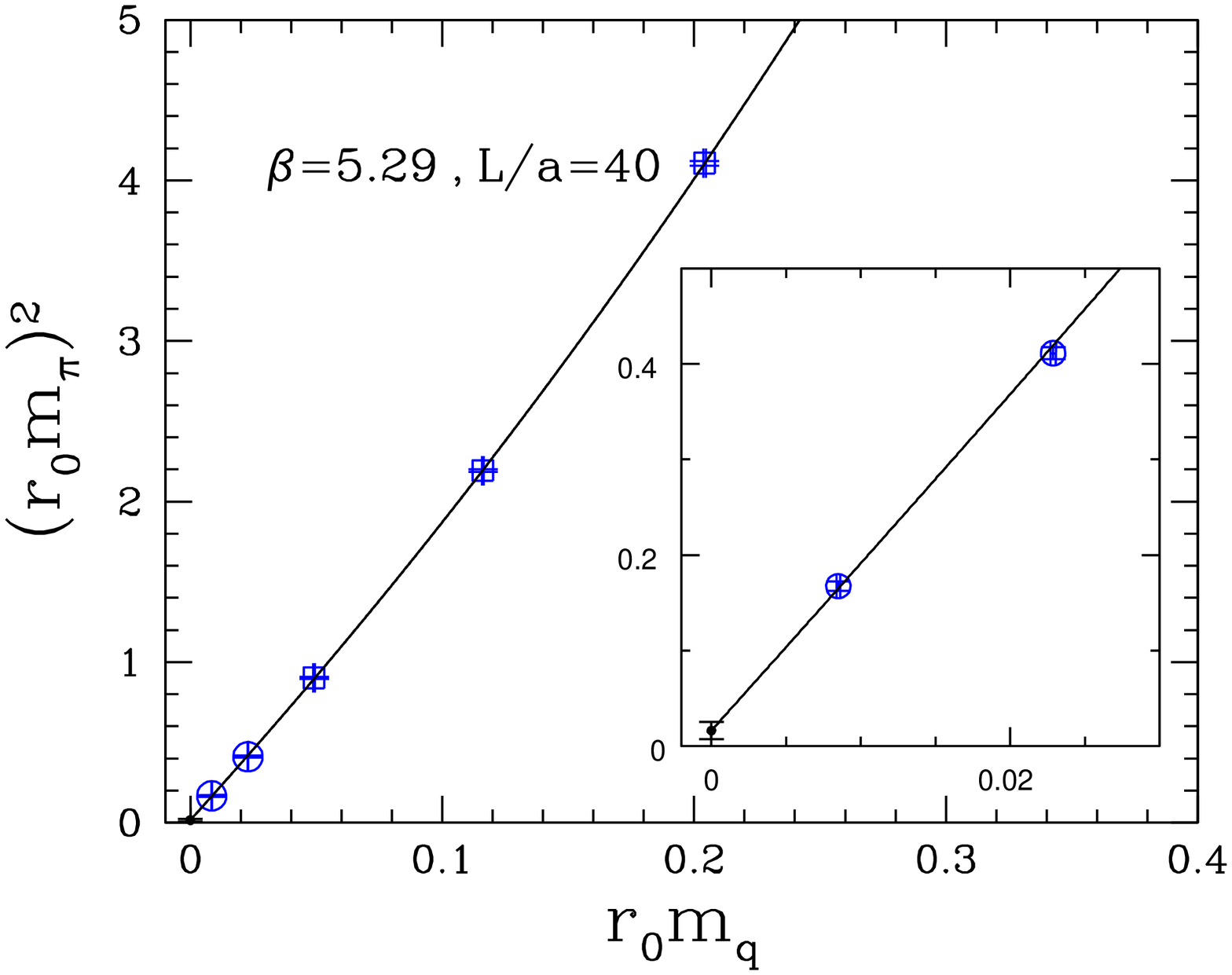,width=10cm,clip=}
\end{center}
\vspace*{-0.4cm}
\caption{The pion mass squared against the quark mass on the $40^3$
  lattice at $\beta=5.29$, together with the chiral fit
  (\ref{ch}). The circles refer to the roman numbers in 
  Table~\ref{tab}, obtained directly on the quoted lattice, while the
  squares are the extrapolated values and
  refer to the italic numbers in Table~\ref{tab}.}
\label{fig40}
\end{figure}

\clearpage
\begin{figure}
\vspace*{-2cm}
\begin{center}
\epsfig{file=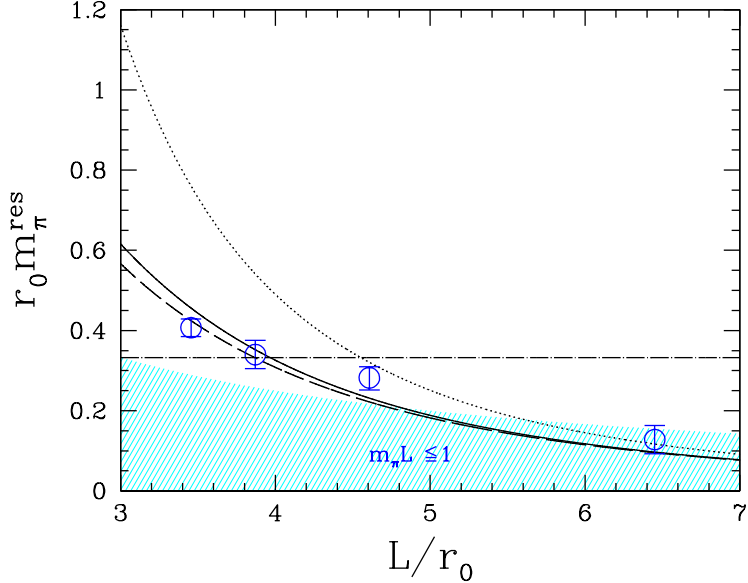,width=10cm,clip=}
\end{center}
\caption{The mass gap as a function of lattice size for
  ($\beta,L/a$)=($5.40,24$), ($5.29,24$) ($5.40,32$) ($5.29,40$), from
  left to right. The
  solid (dashed) [dotted] curve is the prediction of the chiral
  effective theory to NNL (NL) [L] order. The shaded area corresponds
  to $m_\pi L \leq 1$, indicating the $\delta$ regime. The
  dashed-dotted horizontal line marks the position of the physical
  pion mass. We have set 
  $F_\pi=92.4 \, \mbox{MeV}$, and the scale parameters $\Lambda_1$, 
  $\Lambda_2$ have been taken from~\cite{Colangelo:2001df}, as stated in
  eq.~(\ref{lec}).} 
\label{cmres}
\end{figure}

\noindent
while $\bar{l}_3 = 4.2$ is taken from our fit of the pion masses in
Figs.~\ref{fig24a} -- \ref{fig40} (see Table~\ref{res}). 
The extrapolated numbers are printed in italics in Table~\ref{tab}. As
the extrapolation concerns mainly the larger quark masses, any small
error inherent in the extrapolation should not affect our final
result significantly. All masses in question are in the $p$ regime. 

In Figs.~\ref{fig24a} -- \ref{fig40} we plot the pion mass squared against
the quark mass on the $24^3$ lattices at $\beta=5.40$ and $\beta=5.29$, on
the $32^3$ lattice at $\beta=5.40$, and on the $40^3$ lattice at
$\beta=5.29$. 
We fit the data by the formula
\begin{equation}
(r_0m_\pi)^2 = A + B\, r_0m_q\, \left[1 + C\, r_0m_q\,
  \ln(D\,r_0m_q)\right]\, ,
\label{ch}
\end{equation}
which allows for a residual pion mass, $A=(r_0 m_\pi^{\rm res})^2$, in
the chiral limit, and at larger quark masses turns over to the
$O(p^4)$ chiral expansion valid in the $p$
regime~\cite{Colangelo:2001df}. This is to say, eq.~(\ref{ch})  
interpolates between the energy levels in the $\delta$ and the $p$
regime. 

As we are working at fixed $L$, we are actually fitting the
mass dependence of $m_\pi(L)|_L$ rather than $m_\pi$, and one might
expect corrections to the fit formula, which cannot be
accounted for by our ansatz (\ref{ch}). However, this is not
the case. In Fig.~\ref{diff} in the Appendix we compare the leading
finite size correction in the $p$ regime  with the leading
contribution to the mass gap in the $\delta$ regime. At our smallest
quark mass the correction to $m_\pi$
(in  fact to the difference $m_\pi-m_\pi^{\rm res}$) turns out to be a
fraction of the statistical error only for $L/a=40$ and $32$, and
about the size of the statistical error for $L/a=24$. Moreover, the
corrections in the 
$p$ regime are about a factor of $12$ smaller than the mass gap.  

The individual fits are shown in  
Figs.~\ref{fig24a} -- \ref{fig40}. On each of our lattices we obtain a
residual mass, which is distinctly above zero. The results are
summarized in Table~\ref{res}. Barring any finite size effects, the
low-energy constant $\bar{l}_3$ is given by 
\begin{equation}
\bar{l}_3=\ln\left(\frac{\Lambda_3}{m_\pi^{\rm
    phys}}\right)^2\, , \quad  \ln \left(r_0 \Lambda_3\right)^2
=32\,\pi^2\, (r_0 F_\pi)^2\, \frac{C}{B} 
\ln\left(\frac{B}{D}\right) \, .
\end{equation}  
The numbers of the individual fits are listed in Table~\ref{res}. The
average value turns out to be $\bar{l}_3= 4.2(2)$, which lies at the
upper end of the results reported in the literature~\cite{Scholz:2009yz}. 

\begin{table}[t]
\begin{center}
\begin{tabular}{|c|c|c|c|c|c|}\hline
 $\beta$ & $L/a$ & $L$ [fm] & $(r_0 m_\pi^{\rm res})^2$ & $m_\pi^{\rm res}$ [MeV] & $\bar{l}_3$\\
\hline
 5.40 & 24 & 1.61 & 0.166(18) & 172(9)\phantom{0} & 4.05(17)\\
 5.29 & 24 & 1.81 & 0.116(24) & 144(15) & 4.22(32)\\
 5.40 & 32 & 2.15 & 0.080(16) & 119(12) & 4.18(13)\\
 5.29 & 40 & 3.01 & 0.016(9)\phantom{0} & \phantom{0}54(14)
 & 4.26(31)\\ \hline 
\end{tabular}
\end{center}
\caption{The residual pion mass $m_\pi^{\rm res}$ and the low-energy
  constant $\bar{l}_3$.}   
\label{res}
\end{table}

In Fig.~\ref{cmres} we finally compare the residual pion masses of
Table~\ref{res} with the prediction (\ref{rotor}) of the chiral
effective theory. We notice that the chiral series seems to have
converged for $L/r_0 \gtrsim 5$, and below that the NL and NNL order
results differ only by a few percent.  We find good  
agreement with the NNL order formula~\cite{Hasenfratz:2009mp}
for all our lattice volumes, ranging from $L/r_0 \approx 3.5$ to
$\approx 6.5$. This is a bit surprising, as our results are barely
touching the $\delta$ regime. The horizontal line in Fig.~\ref{cmres}
shows the position of the physical pion mass. To capture the physics
of light pions, and the spontaneous breakdown of chiral
symmetry, the spatial extent of the lattice should be $L \gtrsim 3
\,\mbox{fm}$, such that $m_\pi^{\rm res} \ll m_\pi^{\rm phys}$.   

With precise data at (say) $L/r_0 \approx 6$, and quark masses at --
or close to --
the physical point, it should be possible to determine $F_\pi$
accurately. The advantage of such a calculation is that it does not
demand renormalization of the axial vector current. A fit of
(\ref{rotor}) to $m_\pi^{\rm res}$ on our largest lattice, $L/a=40$ at
$\beta=5.29$ (the rightmost data point in 
Fig.~\ref{cmres}), gives the pion decay constant in the chiral limit
\begin{equation}
F \equiv F_\pi|_{m_\pi=0}=78^{+14}_{-10}\,\mbox{MeV} 
\end{equation}
(still using $r_0=0.467\,\mbox{fm}$), which is in the right
ball-park~\cite{Colangelo:2003hf}, but not accurate enough to be
useful yet. 
Due to the relatively small correction of the NNL order
contribution it will, however, not be possible to determine the
low-energy constants $\bar{l}_1$ and $\bar{l}_2$ reliably.  

\section*{Acknowledgement}

The numerical calculations have been performed on the SGI Altix 4700 at LRZ
(Munich), the IBM BlueGeneL and BlueGeneP at NIC (J\"ulich), the
BlueGeneL at EPCC (Edinburgh), and the apeNEXT at NIC/DESY
(Zeuthen). We thank all institutions for their support. Financially,
this work has been supported in part by the EU Integrated Infrastructure
Initiative {\it HadronPhysics2} and by the DFG under contract SFB/TR
55 (Hadron Physics from Lattice QCD). JMZ is supported through the
UK’s STFC Advanced Fellowship Program under contract number ST/F009658/1.

\begin{figure}[t]
\vspace*{-2cm}
\begin{center}
\epsfig{file=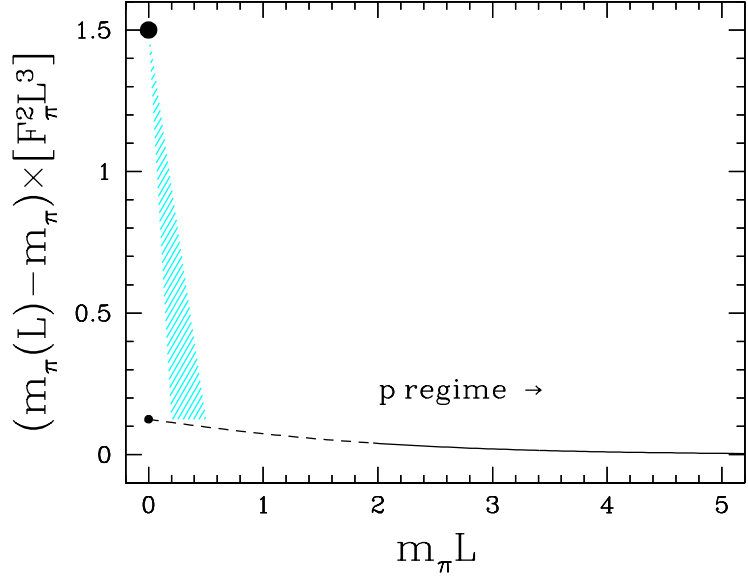,width=10cm,clip=}
\end{center}
\caption{The leading contribution to the finite size mass shift of the pion, in
  units of $(F_\pi^2 L^3)^{-1}$, as a function of $m_\pi L$ (solid and
  dashed curve), compared
  with the leading order contribution to the mass gap (solid circle).
  Within the shaded area $m_\pi(L) \times F_\pi^2L^3 $ drops from
  $3/2$ at $m_\pi L=0$ to $1/8$ at $1/F_\pi^2L^2 \ll m_\pi L \ll 1$.} 
\label{diff}
\end{figure}

\section*{Appendix}

In Fig.~\ref{diff} we show the leading $O(p^2)$ finite size correction
to the mass of the pion, 
\begin{equation}
m_\pi(L)-m_\pi=\frac{m_\pi^3}{16\pi^2 F_\pi^2} \sum_{\vec{n}\neq 0}
\frac{K_1(m_\pi|\vec{n}|L)}{m_\pi|\vec{n}|L} \, ,
\label{lead}
\end{equation}
together with the mass gap (or energy gap) to leading order,
\begin{equation}
m_\pi^{\rm res} = \frac{3}{2 F_\pi^2 L^3} \,.
\label{lead2}
\end{equation}
According to eq.~(25) of~\cite{Leutwyler:1987ak}, the finite volume
shift drops by a factor of $12$ from
$3/2F_\pi^2L^3$ to $1/8F_\pi^2L^3$ in the shaded region. Incidentally
(?), the $p$ regime finite size correction formula (\ref{lead})
extrapolates to exactly the same value, $1/8F_\pi^2L^3$, in the chiral limit.
For small $m_\pi L$ we can express the 
sum over $\vec{n}$ in eq.~(\ref{lead}) by an integral:
\begin{equation}
m_\pi(L)=m_\pi+\frac{m_\pi^3}{16\pi^2 F_\pi^2} \int_0^\infty d\nu\,
4\pi\nu^2 \, \frac{K_1(m_\pi\nu L)}{m_\pi\nu L} \,.
\end{equation}
Changing the variable $\nu$ to $\mu=m_\pi\nu L$, we obtain
\begin{equation}
m_\pi(L)=m_\pi+\frac{1}{4\pi F_\pi^2L^3} \int_0^\infty d\mu\,\mu\,
K_1(\mu)\,.
\end{equation}
The integral is known analytically,
\begin{equation}
\int_0^\infty d\mu\,\mu\,K_1(\mu) = \frac{\pi}{2}\,,
\end{equation}
giving
\begin{equation}
m_\pi(L)=m_\pi+\frac{1}{8F_\pi^2L^3}
\label{lim}
\end{equation}
as the small $m_\pi L$ limit of (\ref{lead}). The $O(p^4)$ contribution to
the mass shift would be suppressed by another factor of $L^{-2}$.

An analytical curve, that smoothly connects the mass gap of the
$\delta$ regime with the $O(p^4)$ correction formula of the $p$ regime, would
be very valuable.

\end{document}